# UNIFICATION AND LOW-ENERGY SUPERSYMMETRY AT ONE AND TWO-LOOP ORDERS

Nir Polonsky

A DISSERTATION
in
Physics

Presented to the Faculties of the University of Pennsylvania in Partial Fulfillment of the Requirements for the Degree of Doctor of Philosophy
1994

**ABSTRACT**

UNIFICATION AND LOW-ENERGY SUPERSYMMETRY
AT ONE AND TWO-LOOP ORDERS

Nir Polonsky

Dissertation Supervisor: Prof. Paul Langacker

The status of coupling constant unification – assuming the validity of the standard model or of its minimal supersymmetric extension at high energies – and of relations between various Yukawa couplings (assuming the supersymmetric extension) which are implied in certain grand-unified theories, are studied in detail. Theoretical uncertainties in the calculations are emphasized, and low-energy constraints and predictions are derived. In particular, we find that bottom-tau unification favors a Higgs boson lighter than 110 GeV. The structure of the vacuum in the model studied is also discussed. Implications of embedding supersymmetric models in grand-unified theories are further explored and are shown to affect the soft supersymmetry breaking mass parameters of the models, and thus the spectrum at low-energy.

## Contents

# Chapter 7

## Conclusions and Future Outlook

In this dissertation we studied in detail various low-energy aspects of supersymmetric theories, and, in particular, of supersymmetric grand-unified theories. We chose the MSSM and the minimal supersymmetric $SU(5)$ and $SO(10)$, all of which are described in chapter 1, as prototype models. Using the data, which is surveyed in chapter 2, we conclude that the relations examined (chapters 3 and 4), i.e.,

- coupling constant unification: $\alpha_1 = \alpha_2 = \alpha_3 = \alpha_G$;
- proton decay (tree level): $M_V \sim M_G \gtrsim 10^{15}$ GeV;
- proton decay (loop level): $M_{H_C} = M_5 \sim M_G \gtrsim 10^{16}$ GeV;
- $SU(5)$ type Yukawa unification: $h_b = h_\tau$;
- minimal-$SO(10)$ type Yukawa unification: $h_t = h_b = h_\tau$;

are all consistent with the data when assuming the MSSM (but not the SM or extensions of either the SM or MSSM with many Higgs doublets) as the effective theory below the GUT scale. The Yukawa unification relations are consistent [assuming $\alpha_s(M_Z) \gtrsim 0.12$] only for large $t$ and $b$-quark Yukawa couplings. The large Yukawa coupling assumption results in additional structure, e.g., the quasi fixed-point convergence, custodial symmetries in the Higgs sector, important GUT effects and new local (sometimes global) minima of the scalar potential along dangerous directions.

If nature is indeed supersymmetric, one has an additional probe of the Planck scale, i.e., the SSB parameters. We have shown that Yukawa unification (i.e., requiring large Yukawa couplings) constrains the SSB parameters space leading (for $\tan\beta \approx 1$) to the upper bound $m_{h^0} \lesssim 110$ GeV on the Higgs boson mass (chapter 5). We also argued that, in general, the SSB parameters do not have universal boundary conditions at $M_G$. This is due to the potentially important GUT renormalization effects (chapter 6). Thus, the soft scalar masses (and not only the SM fermion masses) should bear traces of a grand-unified sector, if it exists. For example, we found [assuming minimal $SU(5)$] $m_{\tilde{t}_{1,2}} \gtrsim 200$ GeV for the $t$-scalars. These GUT effects are enhanced when considering either Yukawa unification or loop-level proton decay (both lead to the large Yukawa coupling assumption).

In our studies we paid special attention to the theoretical uncertainties in the calculations. For gauge and Yukawa couplings we included one-loop threshold and finite corrections due to the Higgs, supersymmetric and superheavy spectra, and from conversion and nonrenormalizable terms, respectively. We also treated $m_t^{pole}$ correlations explicitly. For the Higgs scalar potential parameters and masses, we included important one-loop corrections due to the supersymmetric spectrum and GUT effects. Implications of the latter were studied in more generality. The supersymmetric threshold corrections to the Higgs sector were studied using different methods and we also commented on the two-loop corrections.

We stressed (see also Appendix A) that using the more recent data we predict $\alpha_s(M_Z) \approx 0.13 \pm 0.01$, which is in agreement with $Z$-pole determinations $\alpha_s(M_Z)$, but significantly higher than some low-energy extractions (see Table 2.1). However, due to the GUT-scale threshold uncertainty, even if $\alpha_s \lesssim 0.12$ is known to high precision and our effective parameter $M_{SUSY}$ (which we argued is $\lesssim 300$ GeV) is obtained from observation and accounted for, coupling constant unification will still be consistent with the data, but the predictive power of the models could be altered by requiring correction terms of a specific magnitude and sign. The situation regarding Yukawa couplings is more complicated. A perturbation to the respective GUT relations of the order of the light family Yukawa couplings must exist. In addition, due to the large predicted $\alpha_s$ values, one has to assume $\mathcal{O}(15\%)$ correction terms in order to maintain a successful prediction for the $b$-quark mass. (For large $\tan\beta$ that correction can take the form of finite superpartner loops.) Stronger constraints on $\tan\beta$ are obtained (or alternatively, larger corrections are required) in models with a right-handed neutrino superfield below the GUT scale [172]. This issue, together with the issue of neutrino masses, is left for future studies.

Our studies suggest that the continuation of the investigation of supersymmetric models is well motivated. We believe it should follow a three-fold approach:

1. Continued improvement of analytic expressions and of numerical calculations (e.g., a consistent two-loop calculation of $m_{h^0}$), as well as improving estimations of the theory uncertainties and their model dependence.

2. Study of the soft parameters, i.e., Planck-scale scenarios for the SSB parameters and their low-energy implications.

3. Calculation (within a specific model) of low-energy quantities that are relatively sensitive to the new superpartner and Higgs particles. (That approach could eventually lead to a global analysis of supersymmetric models.)

We chose not to emphasize the issue of non-universality (of the SSB parameters). Obviously, consideration of such scenarios results in compromising the predictive power. It is useful to investigate characteristics of such models (as we did), but the advantage of pursuing such studies any further before any data (regarding the spectrum) is available is not clear.

Lastly, we pointed out (see also Appendix D) the typically complicated structure of the vacuum. The theory contains many scalar fields and the scalar potential could have local (and global) minima in different directions. That observation can be used to constrain the parameter space, as well as have interesting cosmological implications, and it deserves further study.